# Increasing Systemic Resilience to Socioeconomic Challenges: Modeling the Dynamics of Liquidity Flows and Systemic Risks Using Navier–Stokes Equations

**Davit Gondauri,** ORCID: https://orcid.org/0000-0002-9611-3688
PhD, Professor, Doctor of Business Administration, Business & Technology University, Georgia
**Nino Chedia,** ORCID: https://orcid.org/0009-0002-4651-2354
MBA (Finance), Associate Professor, Business & Technology University, Georgia
**Vakhtang Tsintsadze,** ORCID: https://orcid.org/0009-0007-4119-4488
PhD Student, Deputy Minister of Economy and Sustainable Development of Georgia, Georgia

**Corresponding author:** Davit Gondauri, Dgondauri@gmail.com
**Type of manuscript:** research paper

**Abstract:** *Modern economic systems face unprecedented socioeconomic challenges, which make increasing systemic resilience and improving liquidity flow management particularly important. Traditional models (CAPM, VaR, GARCH) often fail to reflect real market fluctuations and extreme events. In the presented study, an innovative mathematical model has been developed, which is based on the interpretation of the Navier-Stokes equations and aims at the quantitative assessment, forecasting, and simulation analysis of liquidity flows and systemic risks. The main hypothesis of the study is that the adapted form of the Navier-Stokes equations in financial modeling allows us to accurately study the internal dynamics of the market, liquidity diffusion, the impact of external shocks, and structural tensions. The model integrates 13 macroeconomic and financial parameters, including liquidity velocity, market pressure, internal stress, beta coefficient, stochastic fluctuations, risk premiums, and contingency factors, all based on real statistical data and formally incorporated into the modified equation. The methodology is based on a mixed approach: econometric testing, Fourier analysis, stochastic simulations, and AI-algorithm tuning, which together provide dynamic testing, calibration, and forecasting capabilities of the model. Simulation-based sensitivity analysis is used, which assesses the impact of parameter changes on the financial balance. The proposed model is empirically validated using macroeconomic and financial data from Georgia for the period 2010–2024. The model is validated using empirical data such as Gross Domestic Product (GDP), inflation, the Gini index, Credit Default Swap (CDS) spreads, and Liquidity Coverage Ratio (LCR) metrics. The results indicate that the model effectively describes the dynamics of liquidity, systemic risks and extreme financial scenarios. The balancing of the model's equations' left and right sides is conducted under real and simulated conditions. When discrepancies occur, a dynamic balance term, represented as a time-varying residual force, ensures systemic adaptation. In addition, the cyclical components obtained by Fourier analysis are harmoniously related to economic cycles and increase the accuracy of the model's predictions. The scientific significance of the study lies in the fact that it creates a mathematically balanced framework for a multifactorial analysis of the behavior of the financial system, which makes it possible not only to predict crises, but also to plan countercyclical policies and increase systemic stability. This model represents a significant scientific advance in the field of economic modeling and creates a real opportunity to increase the resilience of financial systems to socioeconomic fluctuations and systemic risks.*

**Received:** 11.02.2025     **Accepted:** 21.05.2025     **Published:** 04.07.2025

**Funding:** There is no funding for this research.
**Publisher:** Academic Research and Publishing UG (i.G.) (Germany).
**Founder:** Academic Research and Publishing UG (i.G.) (Germany).

**Cite as:** Gondauri, D., Chedia, N., & Tsintsadze, V. (2025). Increasing Systemic Resilience to Socioeconomic Challenges: Modeling the Dynamics of Liquidity Flows and Systemic Risks Using Navier–Stokes Equations. *SocioEconomic Challenges*, *9*(2), 92-113. https://doi.org/10.61093/sec.9(2).92-113.2025

**INTRODUCTION**

The modern world's economic system faces multiple intersecting socioeconomic challenges, the frequency, intensity and global impact of which are constantly increasing. Climate change, geopolitical conflicts, pandemic shocks, financial cybersecurity risks, and technological disruptions create an environment where financial systems are exposed to constant and difficult-to-predict shocks. In such conditions, it is critically important to increase systemic resilience – that is, to create mechanisms that ensure rapid adaptation of markets, maintain liquidity and prevent systemic risks. A sustainable financial system must be not only reactive in times of crisis, but also proactive – in order to identify threats in a timely manner, analyze the dynamics arising from them and be able to provide a strategic response. That is why there is a need to develop analytical and modeling tools that include an integrated analysis of systemic uncertainty, nonlinear changes, and multifactorial connections. Increasing the resilience of financial systems is not just part of the economic agenda today – it is a critical necessity for global stability and equitable development.

In financial systems, liquidity flows and systemic risks play a critical role in the sustainability and development of the economy. Against the backdrop of globalization and the rapid growth of digital financial technologies, financial markets are increasingly dynamic, interconnected, and nonlinear. Historical experience of financial crises shows that liquidity shortages and the strengthening of systemic risks can become the main provoking factors of economic crises.

Liquidity flows reflect the movement of financial resources between economic entities, while systemic risks arise when a financial crisis quickly spreads from one sector to the entire economy. These risks are particularly problematic because they are unpredictable and often cannot be fully assessed by classical financial models.

Liquidity flows and systemic risks represent critical elements in the architecture of modern economic systems. Their comprehensive analysis enables financial authorities to identify early-warning signals of instability, design responsive regulatory frameworks, and enhance overall financial resilience. When the distribution of liquidity is uneven or suddenly decreases, financial institutions find it difficult to carry out their operations effectively, which often leads to banking crises, market panics, and disruptions of the financial system. In addition, systemic risk management is related to the fact that a financial crisis can spread from one sector to another, which further complicates the situation and prevents a rapid recovery of the economy.

The diverse and nonlinear behavior of modern financial systems makes it difficult to accurately assess liquidity flows and systemic risks. Existing traditional models, such as CAPM, VaR, and GARCH, often fail to reflect real market dynamics and fail to predict extreme events. For example, the Capital Asset Pricing Model (CAPM) considers only standard risks and cannot capture nonlinear changes. The Value at Risk (VaR) model is widely used, but it cannot estimate crisis events. The GARCH model is based on statistical distributions, which often fail to describe real market behavior. Given these limitations, there is a need to develop a more in-depth analysis of unstable processes, which is possible with fundamental mathematical approaches.

The methods listed above cannot provide an adequate response to the complex, nonlinear dynamics of the financial system. Accordingly, it is necessary to introduce new, more accurate mathematical tools that will be able to analyze and predict systemic risks in more depth.

Originally designed for modeling fluid dynamics, the Navier-Stokes equations have proven to be a valuable mathematical tool for analyzing financial flows. Their application in economic modeling allows researchers to capture the dynamic, nonlinear nature of capital movement and systemic interdependence within the market. Their integration into financial modeling opens up new possibilities, as they allow for:

- Accurate analysis of the continuity and changes in liquidity flows – The Navier-Stokes equations provide a highly accurate study of the continuity and changes in liquidity flows in financial systems. They describe how financial resources are allocated in markets and how their movement changes under the influence of internal and external factors. This contributes to the development of dynamic modeling of liquidity flows, which allows for a faster response to financial crises.
- Modeling market turbulence and shocks – Financial markets often experience unexpected and severe fluctuations, which can be caused by economic crises, global political and economic changes, or changes in investor expectations. Most traditional financial models fail to accurately predict extreme market movements, while the use of the Navier-Stokes equations allows us to better understand market turbulence, consider the impact of shocks on the market through modeling, and develop appropriate response strategies.
- Developing a new methodology for predicting systemic risks – Early detection and prediction of systemic risks are critically important for the stability of the financial system. The Navier-Stokes equations facilitate the study of structural interconnections in financial networks, which allows us to analyze the dynamics of the spread of systemic risks. This approach is especially important when a crisis spreads from one sector to the entire economy and it is necessary to determine which mechanisms can stop or limit it.
- Improving effective control and management of liquidity flows – Effective liquidity management is one of the most important tasks in ensuring the stability of financial systems. Using the Navier-Stokes equations, it is possible to optimize the distribution of liquidity in such a way as to reduce the risk of financial crises. This approach allows central banks and financial institutions to better control the movement of financial flows and develop more accurate strategies for responding to economic shocks.

The Navier-Stokes equations provide a more accurate response to the complex dynamics of the financial system, which is often difficult or impossible to achieve using traditional models.

The Navier-Stokes equations are also one of the unsolved problems in the field of mathematics. These equations are included in the list of Millennium Problems of the Clay Mathematics Institute, the solution of which is associated with fundamental mathematical and applied challenges. Although these equations are successfully used in engineering and physics to model fluid dynamics, their full mathematical proof and general solution are still open to scientists. The use of this model in financial systems represents an innovative approach that will contribute to a better understanding of financial processes and more accurate forecasting.

The main hypothesis of this study is that the use of Navier-Stokes equations in financial modeling provides more accurate predictions of liquidity flows and systemic risks than traditional models. This approach is based on the idea that financial markets, like fluid flows, contain internal resistances and external influences, which can be mathematically described by differential equations. Such analysis creates a solid basis for crisis prediction and countercyclical policy planning.

The study aims to better study the dynamics of financial systems, liquidity flows and systemic risks using the Navier-Stokes equations. This will help to improve the prediction of economic crises, as well as their prevention and increase the resilience of financial institutions.

- To determine how the Navier-Stokes equations can determine the dynamics of financial flows

The Navier-Stokes equations describe the movement of a fluid and the dynamics of its flows, which makes them an ideal tool for analyzing financial flows. Capital movements in financial markets are often as continuous and variable as fluid flows, which makes the use of this methodology particularly important. The study will examine how these equations can be adapted to financial data and how this approach can help to more accurately model liquidity flows.

- Assessing the effectiveness of systemic risk reduction for financial institutions

Systemic risks pose a threat to the entire economic system, since a single financial shock can spread to the entire sector. Traditional models often fail to describe the dynamics of these processes with sufficient accuracy. Using the Navier-Stokes equations allows us to better understand the interaction of financial institutions and see how risks spread from one to another. The study will assess the accuracy of this method and its effectiveness in reducing systemic risks.

- Developing new strategies for crisis prediction and prevention

Crisis management should not only be focused on their detection, but also on their prevention. Navier-Stokes modeling facilitates the analysis of financial flows in such a way that it becomes possible to detect crisis situations in advance. The research will develop approaches that will help regulators, banks and investors better manage financial flows, prevent systemic disruptions and develop more effective risk reduction mechanisms.

The introduction of this approach into financial systems will significantly improve the management of economic stability and the accuracy of risk analysis. The use of Navier-Stokes equations in financial modeling will create a new paradigm that provides a deeper and more objective analysis of the market, will allow political and financial institutions to identify and manage risks in a timely manner and, as a result, contribute to ensuring financial stability and sustainable economic growth.

**LITERATURE REVIEW**

The analysis of modern financial systems increasingly requires interdisciplinary approaches that combine both economic-financial and mathematical-physical disciplines. In particular, a qualified analysis of the dynamics of market liquidity, the development of systemic risks, and internal structural pressures often goes beyond the capabilities of traditional econometric models. Against this background, there is growing interest in mathematical models inspired by fluid dynamics, one of which is the interpretation of the Navier-Stokes equations in an economic context. The Navier-Stokes equations are traditionally used in physics to calculate the motion of liquids and gases. However, in recent years, various researchers have put forward the idea that analogous dynamical systems can be used to simulate financial phenomena such as the non-uniform movement of liquidity flows, the propagation of shocks in banking networks, or the cascading effects of systemic risks. The financial system, as a complex, nonlinear, and time-varying structure, exhibits fluid dynamics-like characteristics in many ways: asset flows flow across different segments, market pressures can be concentrated or diffused, and disturbances at the micro and macro levels can quickly propagate throughout the system. Accordingly, the use of Navier-Stokes equations opens up a new way to model these complex processes.

Of particular interest in this thematic context is the modeling framework developed by Terence Tao, which not only offers a new interpretive tool but also allows us to study in detail the mechanisms of singularity and instability in dynamical systems. Such an approach directly coincides with the issue of how crisis points are identified in financial systems.

Tao's (2016) study is considered one of the most profound analyses in the context of the Navier-Stokes equations. In this work, the author retains the overall structure of the nonlinear term but adds an average value operator, which gives the equation greater transparency and analytical flexibility. Interestingly, Tao proved in the modeling process that are initial conditions under which the system enters a finite-time singularity (the so-called "blow-up"). This fact, although not directly a final solution to the millennium problem, significantly increases our understanding of the mechanisms that can cause irregularity in real systems.

It is noteworthy that Tao's conclusions indicate that classical approaches may be insufficient to manage the risk of singularity; therefore, an even deeper, model-based analysis is needed. It is precisely such a constructive vision that innovatively develops both theoretical and applied frameworks, especially given that the Navier-Stokes millennium problem is still open to scientists. The intermediate version that Tao uses can itself be considered the space where the mechanism of global regularity breakdown is particularly visible – this approach shows the reader sharply and in many ways what kinds of processes can occur in both mathematical and applied systems.

One of the most fundamental challenges of the Navier-Stokes equations is particularly discussed by Pfefferman (2006), whose work points out the main obstacles related to the existence and smoothness of a solution to this problem. The author describes in detail the formal aspects of the Millennium Prize problem and shows how difficult it is to solve this problem from the point of view of modern mathematical analysis. Pfefferman emphasizes that despite the progress made in cases of partial simplicity in solving the fluid motion model, the difficulties in the full three-dimensional space remain unsolved. In addition, the article pays special attention to the problem of smooth and unique solutions of the time-dependent pressure and velocity fields of a fluid, the existence of which has been proven mainly in limited cases.

In addition, Fefferman discusses the theoretical tools, such as energy inequalities and functional analysis methods, with which current progress in this field is assessed. His research significantly advances the field of mathematical physics and numerical modeling and has a significant impact on modern standards of modeling systemic risks, especially when the use of the Navier-Stokes equations is necessary to describe the dynamics of economic systems.

In terms of new approaches, Sanders et al. (2024) propose a Hamiltonian formulation of the Navier-Stokes equations, which is based on the principle of least action and uses the Hamilton-Jacob equation to provide a new perspective on the analysis of hydrodynamic problems. Within the framework of this innovative approach, it is emphasized that it is possible to introduce completely new mathematical frameworks for solving classical problems.

Similarly, the study by Durmagambetov and Fazilova (2015) deserves special attention for its ambitious claim that the method developed by the authors represents the ultimate solution to the Navier-Stokes equation problem, as indicated by the Clay Institute of the Millennium Challenge Corporation. Their model is based on a time-oriented evaluation of the Cauchy problem and also indicates the possibility that the loss of smoothness in classical solutions is possible under certain conditions. As a result, this work creates a new theoretical framework and provides a significant stimulus for further research.

Among the examples of the use of the Navier-Stokes equations in social and innovative processes, the work of Shinohara and Georgescu (2010) deserves special mention. The model they propose looks at the diffusion of innovation in terms of an analogy to physical systems: the authors compare the spread of innovation to a flow, and diffusion to the process of heat transfer. For mathematical analysis, the Navier-Stokes equations are used to model the movement of settlers – the so-called convection process. Special emphasis is placed on simulating the flow of people in real space, for example, in the Tokyo Aquarium building. The significance of this research goes beyond the analysis of the spread of innovations and is extrapolated to improving the planning of public spaces and evacuation policies.

In recent studies, a special place is occupied by the work of Monyayi et al. (2024), where the conditions for the existence and homogeneity of the fractional Navier-Stokes evolution equation are discussed in detail. The authors use the Banach fixed point theorem and the value of the fractional order $\beta$ of the Atangana-Baleanu-Caputo type in the analysis of the problem, which allows them to study fluid dynamics models from a new perspective. To obtain exact and approximate solutions, the study successfully uses the iterative method, as well as the Laplace and Sumoud transforms, which make it impossible to solve the one-dimensional unsteady flow of a viscous fluid in a pipe. The methodology is based on the representation of both linear and nonlinear fractional differential equations as a series, which significantly simplifies their calculation and contributes to achieving rapid convergence to an exact solution. The results of the study, which are graphically presented using Mathematica software, clearly demonstrate that as the $\beta$-value increases, the solution increasingly approaches the exact value. To illustrate this, approximations of the first four terms are given and the efficiency of their convergence to the final solution is emphasized.

In the field of cash flow modeling and forecasting, an innovative approach is worth mentioning, in which the Navier-Stokes equations are used to study the velocity of money. The dynamic model proposed by Khakousti and Tsiotsios (2022) views cash flow as a turbulent fluid flow, which allows us to analyze macroeconomic variables by analogy with physical parameters. For example, the standard deviation of output is considered as the viscosity of the fluid, changes in interest rates are equivalent to changes in pressure, and financial innovations and institutional factors are represented in the model as the radius of the flow channel. Their approach uses econometric tools – OLS and a two-stage instrumental variables model (IV-2SLS), which is particularly important for reducing the risks of endogeneity, and the predictive power of the model is assessed using RMSD (root-mean-square deviation). The results obtained indicate that the analytical framework based on the Navier-Stokes equations not only enriches the economic interpretation of the determinants of the velocity of money, but also creates a rich foundation for further research in terms of applying fluid flow analogies to financial systems.

One of the most prominent examples of the application of turbulence theory to financial markets is the study by Zeng and Dong (2024). The authors attempted to link the dynamics of financial markets and fluid turbulence, for which they developed an innovative analytical framework that allows them to offer a new interpretation of the complexities of markets. The study focuses on a comparative analysis of several national stock exchange indices and simulations of simplified turbulent velocity time series. Attention is paid to such basic statistical features as probability distributions, correlation structures, and spectral density. The construction of a model of financial market capital flows, the finding of appropriate analytical solutions, and subsequent simulations confirm that turbulence in financial markets does indeed exhibit statistical similarities to physical turbulent phenomena.

However, Zeng and Dong's computer simulations and data analysis also show that the behavior of these two systems is somewhat different. Particularly striking are the different properties of the probability distribution and the time scales of correlations, which confirm that financial markets, despite their statistical similarities, are multi-axis, complex systems and their perfect reflection on natural turbulent events is not possible. Accordingly, the application of turbulence theory in market analysis has its limits and direct transposition does not always work. The paper also reviews the Bezier box technique for simulating market turbulence and various trading strategies developed on the basis of this analysis – they may have practical significance for risk management and financial policy. Ultimately, this research develops a new perspective on econophysics, creating a platform for studying the complexity of markets, financial risks, and strategies related to their prevention.

An innovative approach is also found in the work of Ghosh and Chaudhury (2022), where the Navier-Stokes equations are integrated with modern machine learning methods, including the iForest-BorutaShap module, to forecast stock prices. Their approach builds on the analogy of financial markets and fluid dynamics, while the study focuses on studying the interaction between various economic variables, which significantly improves the accuracy of forecasting stochastic processes. Such an approach contributes to the methodological renewal of stock price valuation systems and may be the basis for the development of new, revolutionary forecasting methods.

Particularly interesting in the field of liquidity risk modeling was the study by Fahim et al. (2025), in which the authors use a dynamic model to study the spread of liquidity risk in banks, and use optimal control theory to assess the effectiveness of central bank interventions. The model is based on data from European banks and simulates various central bank intervention scenarios, thereby showing how systemic financial contagion effects can be reduced under different market conditions. The research results offer practical recommendations for both policymakers and financial institutions to prevent future crises and increase resilience.

The analysis of systemic risk in the banking system and its structural distribution is quite relevant in modern research. For example, May and Arinamipati (2009) draw attention to the fact that banking crises have clearly demonstrated that risk management strategies introduced in recent years often fail to provide effective control of overall systemic risk. Their model analysis shows that the capital reserves, structural characteristics and system of interactions of individual banks create such interdependencies that systemic risks spread not only to a single bank but to the entire financial network.

In terms of detailed dynamic modeling, the work of Fatone and Mariani (2019) is distinguished by the fact that they consider systemic risks in a model of the banking system built on the basis of stochastic differential ratios. The focus here is on the time evolution of banks' logarithmic cash reserves, cooperation, and the dynamics of interactions between banks and regulators. The authors integrate previous models and create new mechanisms for determining the relationship within the system and the connection with the regulator. The system is described as a process of failure probability over specified periods, which is of direct interest to regulators seeking financial stability. Simulation examples show in detail how management effectiveness can be assessed under different shock scenarios – an important contribution to modern standards of financial risk management.

The multifaceted nature of the system's response is particularly emphasized in the study by Allen and Gale (2004), who describe how small, localized shocks can affect the entire financial system. Their model emphasizes that if individual liquidity shocks to banks are not properly regulated, the entire system may reach a single equilibrium based solely on stochastic consumption and asset volatility. These findings are important for predicting asset price volatility and the risk of systemic failures.

As for the clearing mechanism, Eisenberg and Noah (2001) argue that any system with liabilities of firms can be "cleaned" by a clearing payment vector that fairly distributes losses. The authors develop an efficient algorithm that processes information about the systemic risk of each member. Their comparative statistics show that, if non-systemic shocks are present in the system, they sometimes completely reduce the total value of the system or the capital of individual firms.

An example of the use of physics models in exchange rate forecasting is the work of Cartono et al. (2020). Here, numerical solutions of the Navier-Stokes equations are used to explain the turbulent flow of the exchange rate in the face of changes in economic parameters. Their analogy implies that, for example, GDP, trade balance, and inflation act as physical parameters, and the forecast is carried out using the Crank-Nicholson method - the one-year forecast almost completely coincides with the actual data.

A more recent approach to modeling and forecasting economic processes is proposed by Briet (2024), who combines the Navier-Stokes equations with the Hamiltonian principle of action and artificial intelligence. In the "Acceleration Balance Equation" (ABE) model, economic processes are considered as geodesic trajectories in the potential field space, which connects their evolution to the laws of natural dynamics and reduces the need for human intervention. This model is integrated into the structure of a neural network, which significantly increases the accuracy of forecasts in both the short and long term.

Duffy (2015) reviews the use of Partial Differential Equations (PDEs) in financial mathematics, where numerical solutions provide a basis for risk analysis of market-traded instruments. The author identifies time-dependent convection-diffusion-reaction PDEs that effectively characterize the dynamics of financial products – including stocks, options, commodities, and derivatives. The historical role of the finite difference method and Black-Scholes-type equations in financial modeling has been clearly demonstrated, which is the basis for the progress achieved in computational finance in recent years.

In the work of DuCourno (2021), the stock market is presented as a physical system with fluid dynamics, where the supply and demand of market agents create forces that have a decisive influence on the behavior of the entire system. The author tries to prove that the dynamics of the stock market can be described by Stokes' law, where specific characteristics of the market, such as viscosity and density, can be directly analogized to the properties of the fluid. The same article focuses on the use of the Reynolds number to classify market conditions – that is, to determine whether the market is laminar, transient, or chaotic, and the analogy of the Mod diagram provides a physical basis for assessing the risks of assets and indices. Computer simulations allowed the author to confirm that Stokes' law does indeed effectively describe market behavior.

In the field of financial analysis and corporate valuation, an important place is occupied by the research of Damodaran (2025), where the Equity Risk Premium (ERP) is considered as the price that the market assigns to the risk of an investment. Damodaran comprehensively analyzes the economic factors that determine ERP – here the perception of investment risks, imperfect market information and macroeconomic expectations are considered. The author is not satisfied with only the historical method of the difference in annual returns between stocks and bonds and considers alternative approaches to both survey and implicit valuation. The survey method is based directly on the opinions of managers and investors, while the implicit approach extracts ERP from the expectations contained in market prices. The article focuses on analyzing the relationship between ERP, default spreads, and real estate capitalization ratios, which makes it possible to differentiate the expected risk premium across sectors. Finally, Damodaran evaluates the differences in the values obtained by these different methods and determines how to choose the right approach to using ERP in different situations.

A distinctly different view is offered by Mausbusin and Callahan (2024), who focus on the popularity of market multiples in corporate valuation practice and the impact of these multiples on market and investor perceptions. The authors rely on two concepts of Damodaran in their analysis: "price formation" and "valuation." Mausbusin and Callahan emphasize that most investors spend most of their time on pricing – choosing the appropriate multiple for the indicator – while valuation requires in-depth fundamental analysis and discounting future free cash flows. According to the survey results, 93% of more than 2,000 CFAs use market multiples in practice (mainly P/E and EV/EBITDA), although 79% also use the discounted cash flow (DCF) model, but in the end, even in this model, there is often a return to the use of multiples. Thus, the DCF model in most cases turns into a form of analysis of multiples. This trend indicates the dominant role of multiples in valuation practice and also shows how the process of "pricing" will eventually grow into the structure of fundamental analysis.

A significant contribution to the analysis of the impact of economic uncertainty on expected stock market returns is made by Bollerslev et al. (2008). The authors use a general equilibrium independent model to examine how the time-varying variance risk premium captures market volatility. The central finding is that the difference between implied and realized variance – or the variance risk premium – effectively explains a significant portion of the time-varying returns on US stock markets since 1990. The study emphasizes that a high variance risk premium is indicative of higher future returns, while a low premium indicates more modest future returns. The mathematical framework is based on the concept of model-independent implied variance, which distinguishes their approach from the traditional Black–Scholes framework. Realized variance is calculated using high-frequency, daily internal data, which significantly increases the accuracy of the forecast. The authors show that the variance risk premium over a quarterly horizon significantly outperforms such common variables as the P/E ratio, the default spread, and the consumption-volume (CAY) index in predictive power. Their findings are particularly important for investment strategies that focus on time-based market valuation and volatility analysis.

Tail risk analysis based on sharp declines in fund prices is developed by Kelly and Gian (2013). Their approach allows us to observe changes in systematic tail risk over time without using options market data, which significantly simplifies the analysis and increases the universality of the model. The resulting measure is closely related to the tail risk based on S&P 500 options, but the authors note that their approach is even more robust over the long term. The study found that increasing tail risk correlates with increasing future excess market returns, and companies with high tail risk tend to have high alpha. These factors point to the critical role of tail risk in determining asset prices and investment decisions.

A more complex aspect of the impact of macroeconomic uncertainty is explored in Segal and Shalyastovich (2023). Their analysis reveals an unexpected paradox: while economic uncertainty typically reduces investment and capital appreciation, it also increases physical capital accumulation. This seemingly contradictory result is explained by the fact that under conditions of uncertainty, companies reduce the level of depreciation and use of capital, which ultimately exceeds the effect of reducing investment. In the model, firms respond to uncertainty not by increasing investment, but by reducing the intensity of capital use, which is a new mechanism of preventive accumulation. It is noteworthy that, according to the model, uncertainty shocks account for about 25% of the formation of stock premiums, and flexible adjustment of capital use allows us to better explain the sensitivity of industrial sectors to uncertainty.

The work of Hamble and Soimark (2018) is a new word in the modeling of systemic financial risks –the framework based on Stochastic Partial Differential Equations (SPDE) they present aims to deeply analyze the effects of endogenous contagion. The innovativeness of the model lies in the fact that it originates from the existing diffusion system, where the default distances of financial institutions are assumed to have a zero intercept, which ensures accurate qualification of crisis events. The system integrates a common noise source and a mean reversion drift component, which are associated with the individual fluctuations of each institution. It is also important that the model takes into account the mechanism of endogenous contagion: the default of one institution can exacerbate systemic risk in other institutions – this is precisely what reflects the nature of real contagion effects in the network. This process is precisely described by the unfavourable SPDE equation, which defines a specific McKinney-Walso diffusion. Hamble and Soimark (2018) innovatively introduce a boundary Dirichlet bridge-type estimation method, which allows for the simulation of large clusters of systemic defaults; depending on the overall noise and mean reversion parameters, the SPDE model often leads to rapid mass loss, which emphasizes the risk of crisis.

A modern approach to the network dynamics of credit institutions can be found in the work of Capone et al. (2020). The model they developed describes in detail the solvency and liquidity processes of banks: each bank is defined by initial reserve requirements and has an independent risk profile, which is reflected in the system by Brownian motion. The interactions between banks present a hierarchical cascade structure, where transactions – borrowing, issuing, or repaying processes – reflect clustered financial activity. An important methodological innovation is the identification of a weak threshold for these processes as the number of banks increases, which allows for the asymptotic estimation of two key macrometrics: the liquidity stress index and the concentration index. These, in turn, serve as key indicators of systemic risk dynamics.

The interplay between fundamental shocks, liquidity risk, and feedback mechanisms is assessed by Kapadia et al. (2013). The authors develop a "danger zone" approach that models in detail how simultaneous shocks affect banks' funding liquidity. Integrating this approach with flexible response strategies allows us to assess the speed and extent to which a liquidity crisis can spread throughout the financial system. In times of crisis, the model will be used to generate the RAMSI (Risk Assessment Model for Systemic Institutions) model, which shows how the combination of funding and spillover effects exacerbates systemic risks and how financial stress can worsen in the event of structural disruptions.

The concept of the Systemic Liquidity Risk Index (SLRI) is significantly developed by Severo (2012), whose work examines in detail how this index affects the equity returns of global banks. The author analyzes data from 53 large banks and concludes that there is no direct correlation between equity returns and SLRI, although at critical moments when liquidity conditions become tight, banks' cash turnover increases significantly. It is noteworthy that there is no clear relationship between bank size and sensitivity to the impact of SLRI, but sensitivity to systemic liquidity risk is positively associated with the Net Stable Funding Ratio (NSFR). Based on this, Severo determines the value of liquidity support from the state and what the insurance premium is that individual banks may be required to cover social costs.

The dynamics of systemic risk in banking ecosystems are reviewed by Irakos et al. (2023). Their model pays special attention to the relationship between troubled and healthy banks, exploring the conditions of equilibrium and stability under different scenarios. The equilibrium points obtained by solving the equations are evaluated in both homogeneous and heterogeneous conditions, and numerical simulations clearly show the features of the dynamics of systemic risk. The theoretical results shed light on how troubled banks operate in different environments, which significantly strengthens the policymaking toolbox in the banking sector.

The theoretical basis for modeling financial crises and default cascades is presented in the work of Minka and Amin (2012). Their analysis became especially relevant in light of the experience of the 2007 financial crisis, when hundreds of banks failed. The authors describe the mechanism of economic shocks that may start with the insolvency of only a few institutions but due to complex financial linkages, lead to a large-scale systemic cascade and widespread defaults. The model they propose combines detailed balance sheet analysis, and jointly considers the development of both insolvency and non-monetary cascades in the financial network. The paper contains asymptotic results on the size of the default cascade and discusses how such effects can propagate and generalize across different financial channels and economic conditions.

Damodaran (2003) extends the traditional perception of risk with an interesting perspective. The author points out that risk can become not only a threat to a company but also a source of competitive advantage, if it is correctly perceived and managed. While classical approaches to risk management focus only on risk reduction – mainly through the use of hedging instruments – Damodaran expands this framework and speaks about the possibility of using risk strategically. The author notes that most analysts limit the assessment of risk to market risk and reflect it only through discount rates, although in reality risk management can enhance the potential of financial flows, redirect investment policy and contribute to the formation of sustainable competitive advantage. The study also offers practical steps – it provides recommendations for companies to develop a risk management strategy that goes beyond hedging risk but also uses it in new ways to increase growth rates and excess profits.

A follow-up paper by Damodaran (2006) focuses on the impact of financial stress on traditional valuation methods – whether discounted cash flow (DCF) models or comparative analysis. The author highlights the problem that stress is often completely ignored or superficially considered, which can lead to a valuation that does not reflect the true state of a company in terms of its ability to meet its obligations. The paper examines in depth how well DCF models capture financial stress, when they ignore this effect, and how the impact of stress can be excluded or included in the analysis. In the final section, Damodaran examines the effect of financial stress on comparative valuations and discusses practical ways in which this impact can be integrated into comparative pricing analysis, making it a more accurate analytical tool for companies.

Although there is significant and interesting research on modeling the dynamics of liquidity flows and systemic risks in financial systems, the topic of using Navier-Stokes equations to study these processes has not yet been widely explored. Thus, our study represents an innovative contribution, because first of all, we propose the use of Navier-Stokes equations to accurately link liquidity flows and systemic risks, which represents a new approach in this field.

**METHODOLOGY**

The increasing complexity and nonlinear behavior of modern financial systems require the implementation of models that simultaneously take into account time dynamics, structural constraints, and external shocks. The basis of this study is the use of Navier-Stokes equations for the analysis of financial flows, which is based on the mathematical analogy of fluid dynamics and integrated simulation modeling of economic parameters.

The methodology is based on a mixed approach that combines both quantitative analysis and theoretical modeling:
- Quantitative analysis includes statistical processing of macroeconomic data, econometric testing, correlation and volatility analysis, modeling of liquidity, risk, Gini coefficient, inflation, and other indicators;
- Theoretical modeling is based on a nonlinear model created by the economic interpretation of the Navier-Stokes equations, which integrates time-varying differentiations, stochastic processes, and Fourier analysis.

**Methodological Foundations**

In the context of financial systems, the Navier-Stokes equations are considered as a dynamic modeling tool that describes the movement of capital flows, the spread of systemic risks, and market reactions to external shocks. Based on mathematical analogy, these equations are transferred to economic models and integrated with parameters such as liquidity velocity, market pressure, and internal resistances. As a result, a basis is created for the analysis of complex phenomena that cannot be covered by traditional models.

The Navier-Stokes equations are used in physics to model fluid and gas dynamics, but they can also effectively describe financial flows, since economic processes often exhibit behavior similar to fluid flows. In this study, the application of the Navier-Stokes equations in financial systems allows us to understand the dynamics of liquidity movement, internal market resistances and turnover factors, the spread of systemic risk in financial instruments, , and the mechanisms of crisis emergence and spread.

*Mathematical Model and Financial Interpretation*

The general form of the Navier-Stokes equations is:

$$\rho(\partial_t \partial v + (v \cdot \nabla)v) = -\nabla P + \mu \nabla^2 v + F \tag{1}$$

where, ρ denotes the inertia or mass of the financial system, which reflects the market's resistance to change and its ability to absorb shocks; v represents the liquidity velocity, which reflects how quickly capital circulates in the economy; ∇P denotes the market pressure gradient, which reflects the direction of pressure on prices in the face of macroeconomic imbalances; μ∇²v represents internal market frictions or liquidity diffusion, which reflects the delays in the movement of capital; F represents external economic pressures, such as central bank policy, GDP cycles, or global financial shocks.

*Impact of Navier-Stokes Equations on Financial Modeling*

1. The model enables forward-looking simulations of liquidity fluctuations under varying economic conditions, thereby allowing early identification of shortages and policy response calibration.

2. By applying the Navier–Stokes equations, the model allows for in-depth evaluation of market cyclicality and periodic fluctuations of financial flows across different economic phases. It offers a dynamic alternative to conventional cycle models by capturing amplitude and frequency variability.

3. The model supports the dynamic evaluation of systemic risk by integrating internal market stress factors and external shocks, thereby enhancing the precision of systemic stability diagnostics.

*Integrating Systemic Risk with the CAPM Model*

Systemic risk is integrated into financial flows through stochastic modeling using the CAPM (Capital Pricing Model). In this context, the beta coefficient ($\beta$) determines the sensitivity of the asset to the market:

$$\beta_i (r_m - r_f) \tag{2}$$

where $\beta_i$ is the asset's sensitivity to the market; $r_m$ is the expected market return; $r_f$ is the risk-free rate.

*Stochastic Processes and Market Volatility*

Stochastic processes represent random fluctuations within the financial system that contribute to short-term volatility and shock transmission. In the presented model, this randomness is formally expressed through the σWt term, which captures exogenous shocks such as geopolitical tensions, abrupt inflation spikes, or credit events. These dynamics introduce necessary uncertainty to test system stability

For this purpose, the Wiener process is used σWt, where $\sigma$ represents the magnitude of market fluctuations.

To identify periodic trends in economic indicators, both trigonometric modeling and Fourier series decomposition are employed. The trigonometric representation uses sine and cosine functions to simulate oscillations in key parameters. In parallel, the Fourier series expansion of v(t) enables the model to extract dominant economic cycles by combining multiple harmonic components, each representing distinct frequencies in macroeconomic behavior.

1. Trigonometric model

A simple trigonometric function is often used for time-cycle analysis, which is given as follows:

$$A \sin(\omega t + \varphi) + B \cos(\omega t + \varphi), \tag{3}$$

where, A and B are amplitude components, ω is the angular frequency, t is time, and φ is the phase shift.

2. Fourier series

A Fourier series representation is often used to study the cyclical components of financial time series:

$$v(t) = \sum A_n \cos(n\omega t) + B_n \sin(n\omega t), \tag{4}$$

where, $A_n$ and $B_n$ are the amplitudes of each harmonic, n is the harmonic number (n = 1, 2, ..., N), $\omega$ is the fundamental frequency, and t is time.

Unified mathematical model
All the above components are integrated into a single algebraic equation:

$$\sigma W_t + \rho\,(\partial v/\partial t + (v \cdot \nabla)v + F_{(GDP)}(t)) = -\nabla P + \mu \nabla^2 v + \eta(r_m - r_f) + CDS + \beta + \varepsilon \qquad (5)$$

where, $\sigma W_t$ is the stochastic shock component; $F_{(GDP)}(t)$ is the GDP-driven cyclical force, $\eta(r_m - r_f)$ is the market risk premium, and CDS, and $\beta$, and $\varepsilon$ are the credit risk, systemic risk, and other compensation factors, respectively.

This equation integrates key structural and stochastic components affecting financial systems. The left-hand side reflects internal system dynamics and external drivers. The term $\sigma W_t$ captures stochastic shocks stemming from unpredictable macroeconomic or geopolitical events. The product $\rho(\partial v/\partial t + (v\,\nabla)v)$ represents systemic inertia and internal stress propagation, while $F_{(GDP)}(t)$ introduces GDP-driven cyclical economic forces. The right-hand side includes suppressive and compensatory components: $\nabla P$ represents macroeconomic pressure, such as inflationary shocks; $\mu \nabla^2 v$ denotes liquidity diffusion hindered by institutional frictions; $\eta(r_m - r_f)$ expresses the effect of market risk premia on liquidity; CDS and $\beta$ are incorporated to account for credit risk and sensitivity to cyclical volatility; and $\varepsilon$ serves as a residual term absorbing unidentified influences not captured elsewhere in the model.

*Simulation Scenarios on the Navier-Stokes Model*

The simulation scenarios implemented within the Navier–Stokes framework utilize real and hypothetical stress conditions to test liquidity resilience and systemic risk amplification. The scientific significance of this approach lies in its ability to translate physical fluid dynamics into financial stress behavior, providing a novel lens for evaluating economic fragility. From a practical standpoint, this model informs policymakers and financial regulators about potential intervention strategies under high-risk conditions.

*Method used.* This analysis uses multiparameter simulation-based sensitivity modeling (simulation-based sensitivity analysis), which involves studying the impact of multiple parameters on the economic version of the Navier-Stokes equation simultaneously or in stages.

This approach is based on the following steps:
- Balancing the initial equation of the Navier-Stokes equation in the baseline scenario (baseline condition);
- Targeted change of important parameters (e.g., inflation, banking pressure, stochastic shock, beta, etc.);
- New calculation of LHS and RHS components for each change;
- Estimation of imbalance (RHS − LHS) and, if necessary, revision of the balance ($\varepsilon$);
- Determination of the degree of sensitivity –which parameter has a high impact on the system balance;
- Conclusion on the need for compensation and recommendation on whether fixed $\varepsilon$ is sufficient or should be transformed into a dynamic form $\varepsilon(t)$.

*Scientific Significance and Application.* The given method represents a modern approach to studying systemic stability, which is used both in macroeconomic models (e.g., DSGE, stochastic models), and in policy analysis, stress testing, and financial forecasting. Through it, is possible to identify critical points in the system, simulate the effects of shocks in real conditions, facilitate countercyclical policy planning, enable automatic response mechanisms in the model (adaptive $\varepsilon(t)$).

*Methodological structure of the study.*

To confirm the accuracy and reliability of the model, an empirical analysis was conducted based on historical data on macroeconomic indicators of the Georgian financial system. The study examined correlation and regression relationships between indicators, which helped to determine and diagnose the internal structure of the model in detail. In parallel, simulation testing was carried out under various stress scenarios, including liquidity crises, inflationary shocks, increased market pressure, and increased systemic risk. The model also included mechanisms for conditions tailored to real and hypothetical circumstances. The expansion of the model also includes artificial intelligence and deep learning methods, which makes it possible to identify nonlinear relationships between variables that cannot be captured by traditional econometric models. Fourier series were used to separate economic cycles, and machine learning algorithms were adapted to pre-estimate temporal parameters. This integration increases the flexibility and predictive power of the model in complex economic conditions. The model also provides for the embedding of real-time operational data from financial markets, which enhances its adaptability and the possibility of periodic verification. The combination of a meta-econometric framework and artificial intelligence forecasting tools significantly enhances the predictive accuracy of the model and its responsiveness to macroeconomic changes. The presented model creates a new platform for assessing the dynamics of liquidity flows in cyclical and nonlinear economic environments. The model allows for the quantitative assessment of systemic risks and the creation of optimal strategies for their reduction. This improves the forecast of financial stability and provides a more informed basis for policy formation.

**RESULTS**

The presented results are based on empirical analysis and simulation modeling of the Georgian financial system. The estimation of all parameters and the simulation results reflect the characteristics that characterize the Georgian economy – including moderate liquidity velocity, high sensitivity to external shocks, and the dominance of the banking sector in credit distribution. These factors directly affect how liquidity flows and systemic risks are formed in both stable and crisis environments. The model is based on statistical data from the National Bank of Georgia and Geostat, which ensure that the formula based on the Navier-Stokes equation realistically reflects the dynamics of the Georgian economy in 2010–2024. Accordingly, the presented results should be analyzed in the context of the structural features of the Georgian economy, the policy environment, and its relationship to regional and global volatility. The analysis is based on both empirical statistical data (e.g., velocity of money, inflation, LCR) and theoretically calibrated parameters (e.g., Fourier analysis, risk premium, β). Each component is integrated into a modified version of the Navier-Stokes model, which is used to describe the behavior of the financial system in a multifactorial manner.

Table 1 presents the set of macroeconomic and financial parameters used for simulation purposes. All input values are derived from empirical data specific to Georgia during the period 2010–2024. The main data sources include the National Bank of Georgia, the National Statistics Office (GeoStat), IMF macroeconomic indicators, and internationally recognized market datasets such as the S&P Emerging BMI Index, J.P. Morgan EMBI, and Damodaran's sectoral risk metrics. These empirically grounded parameters were applied to test the behavior of liquidity flows, systemic vulnerabilities, and financial frictions within the adapted Navier–Stokes equation.

**Table 1. Summary of Parameters Used in the Modified Navier-Stokes Model**

| Parameter Title | Result / Interpretation |
|---|---|
| Relative volatility analysis – the basis for calculating market inertia | $\rho = 1 / 1.35 \approx 0.7421$ – indicates the relatively low sensitivity of the emerging market. |
| Internal tension – $(v \cdot \nabla)v$ and Gini | $(v \cdot \nabla)v = 0.36$ – moderate inequality within the market, based on Gini. |
| Change in liquidity velocity $\partial v / \partial t$ | $\partial v / \partial t = -0.23$ – money circulation is decreasing, negative momentum in the market. |
| Money circulation speed – v | $v = 3.46$ – moderate money circulation speed, moderately active economy. |
| Market pressure (PPP) | Inflation ≈ 1.9% – low demand-supply pressure, stable price formation. |
| Market stickiness – μ | $\mu = 66.06\%$ – high banking pressure, disruption of liquidity movement. |
| Systematic risk – beta | $\beta = 1.2$ – typical sensitivity of an emerging market to external shocks. |
| Liquidity diffusion – $\nabla^2 v$ and LCR | $\nabla^2 v = 100\%$ – banks' liquidity is ensured during stressful periods. |
| Stochastic shock – $\sigma W_t$ | $\sigma W_t = 1.3475$ – there are moderate stochastic fluctuations in the market. |
| Market risk premium – $\eta(r_m - r_f)$ | $\eta(r_m - r_f) = 8.35\%$ – expected return on risky assets. |
| Fourier analysis of GDP | $F_{(GDP)}(t) = 0.07577$ – cyclical external force on economic waves. |
| Credit default swap – CDS | CDS = 2.98% – medium level, moderate default risk interaction. |
| Unintended economic effects – ε | $\varepsilon = 0.4547$ – a set of components that are formally unquantifiable but have a significant impact on market behavior. |

*Source: authors' calculations based on data from S&P Dow Jones Indices (2024), J.P. Morgan (2024), National Bank of Georgia (2024), Damodaran (2025), and IMF (2024).*

The relative volatility method was used to determine market inertia (ρ). The following two financial indices were compared: the S&P Emerging BMI Index, an index of market capitalization of emerging markets, and the iShares J.P. Morgan USD Emerging Markets Bond ETF (EMB), which reflects government bonds of developing countries denominated in dollars.

For each index, a time series of prices was collected with daily data from January 1, 2020, to December 31, 2024. From the data obtained, standard deviations were calculated annually: Std Dev (BMI) and Std Dev (JPM Sov Bond). Annual averages of standard deviations were calculated for a five-year period. The obtained averages were compared to calculate relative volatility, using the formula:

Relative Volatility = Avg[Std Dev (BMI)] / Avg[Std Dev (JPM Sov Bond)]  (6)

The obtained value was 1.35. Therefore, $\rho$ = 1 / Relative Volatility = 1 / 1.35 ≈ 0.7421

*Interpretation:*
This indicator indicates that the volatility of the emerging market equity segment (BMI Index) is 1.35 times higher than that of sovereign bonds (EMB).

*Market inertia ($\rho$):*
To estimate market stability in the Navier-Stokes equation, the inverse relationship is used:

$\rho$ = 1 / Relative Volatility = 1 / 1.35 ≈ 0.7421.  (7)

This means that the market has relatively low sensitivity – high inertia – which is important for estimating the stability of the financial system in a dynamic model.

The model $(v \cdot \nabla)v$ in the Navier-Stokes equation captures the effect of internal movement in the system – that is, how the existing liquidity vector affects other directions. In an economic context, it can appear as internal structural tension, which is caused by inequality. The "Gini coefficient", which is an index of income or wealth inequality in a society, is used as a quantitative indicator of this structural imbalance. It is 0.36. This value indicates that there is a moderate level of inequality within the economic system, which causes internal tension – weak or uneven movement of liquidity in different segments of the economy.

In the Navier-Stokes equation, the model $(v \cdot \nabla)v$ is introduced as a force causing uneven distribution. Its quantitative notation, based on the Gini coefficient, describes how liquidity is dispersed within the economy. The higher the Gini, the more unbalanced the distribution of resources becomes. Accordingly, the tension expressed in this term should be reflected in the model as a positive impact in those regions where liquidity accumulates, and negative where it is deficient.

Change in liquidity velocity $\partial v/\partial t$ expresses the change in the velocity of money circulation over time. It is the dynamic component of the Navier-Stokes equation and describes how the movement of liquidity in the market changes.

In this case, it was calculated as follows. In 2023, the liquidity velocity was 3.69. In 2024, the liquidity velocity decreased 3.46.

*Calculation:*

$\partial v/\partial t$ = 3.46 - 3.69 = -0.23  (8)

*Interpretation:*
A negative value indicates that the money supply in the economy has slowed down. This change may reflect a decrease in demand, economic inertia, a slowdown in growth or a deterioration in expectations.

*Role of the model:*

In the Navier-Stokes equation, $\partial v/\partial t$ reflects the change in market dynamics over time. Its inclusion gives a realistic picture to the model and expresses the momentum that either accelerates or slows down liquidity flows.

A value of -0.23 indicates short-term dynamic restraint, which is an important signal for systemic assessment.

The velocity of money (v) expresses the number of times money is used in the economy in a year. It is one of the main indicators for assessing liquidity dynamics and is used as the internal velocity component of the Navier-Stokes equation. The velocity of money in 2024 is v = 3.46. This means that money circulates in the economy on average 3.46 times per calendar year.

*Interpretation:*

A high velocity of money is associated with increased economic activity, increased demand, and increased fiscal and monetary dynamics.

*However:*

v = 3.46 is considered below average in the global context, indicating money moves relatively slowly in the system, consumers or businesses are saving and spending less, and the level of investment may not be high.

In the Navier-Stokes model, v represents the velocity, to which internal tension ($\nabla v$) and time change ($\partial v/\partial t$) are added. In the model, v = 3.46 acts as the core of the basic liquid dynamics and is used in combination with diffusion and pressure phenomena.

Market pressure (PPP) reflects the economic tension that arises from the interaction of demand and supply. Inflation is one of the most visible manifestations of this pressure. The annual inflation rate in 2024 was 1.9%. This data indicates low supply-demand pressure and a controlled level of inflation.

*Interpretation:*

A low inflation rate of 1.9% indicates that consumer demand may be reduced, the supply side is functioning stably, economic activity is moderate, and monetary policy is effective and performs a stabilizing function.

In the Navier-Stokes model, $\nabla P$ represents the gradient of inflationary pressure in the market –more simply, it is the price pressure on the system. Low inflation means that the market does not experience significant price fluctuations and does not need high liquidity to respond.

In the economic adaptation of the Navier-Stokes framework, the coefficient $\mu$ is interpreted as a measure of internal market stickiness, capturing the extent to which financial flows are constrained by institutional or structural rigidities. In this study, $\mu$ is proxied by the level of domestic credit to the private sector as a percentage of GDP, which stands at 66.06% in the case of Georgia. This value suggests that a substantial portion of financial resources is channeled through the banking sector, reflecting a relatively rigid allocation structure.

A high $\mu$ indicates that financial shocks are less likely to be transmitted rapidly, as liquidity tends to be retained within institutional frameworks, delaying response times to economic changes. Conversely, a low $\mu$ would imply a more fluid market where capital reallocates quickly in reaction to shocks. In the Navier-Stokes equation, this parameter appears in the diffusion term $\mu^2 v$, governing how liquidity disperses across the economic domain. The elevated stickiness observed in this context implies that despite adequate liquidity, the flow dynamics remain constrained, contributing to potential mismatches between monetary policy signals and actual market behavior.

Systemic risk (beta, $\beta$) represents the overall sensitivity of the market to external shocks. In financial modeling, it is used to measure the extent to which a financial system is susceptible to broader market volatility. $\beta = 1.2$ is taken from the leveraged beta data from A. Damodaran's model, which reflects the typical systemic risk of emerging markets. Interpretation:

A beta of 1.2 means that the Georgian financial system is above average sensitive to external economic changes; any global or regional crisis may "be amplified" in the Georgian market; the market overreacts to economic expectations and investment fluctuations. For emerging economies, this beta estimate is considered a realistically high-risk segment. Based on Damodaran's data, a leveraged beta of 1.2 reflects the impact of capital structure and the level of market volatility.

In the Navier-Stokes model, beta can be written as a systemic risk term, which is an internal indicator of the structural instability of the market.

This model can directly affect the system's disequilibrium or add to the risk premium.

In the Navier-Stokes model, $\nabla^2 v$ expresses the intensity of the movement of fluid (or in our case – liquidity) in space. For the economic model, this term reflects how quickly and efficiently liquidity moves within the financial system – especially in the banking sector. In Georgia, the liquidity coverage ratio (LCR) is defined as: total and foreign currency LCR: 100%. LCR of GEL: 75% For the model, it is assumed:

$\nabla^2 v = 100.00\%$ (based on the total LCR).

*Interpretation:*

The purpose of the LCR is to ensure that banks have sufficient liquidity to cover expected cash outflows during a 30-day stress period. A requirement of 100% indicates that the Georgian banking sector is required to have sufficient liquid resources, which is an indicator of the stability of the system.

$\nabla^2 v$ acts as a liquidity dispersion mechanism – a high value indicates that liquidity is evenly distributed in the system and there are no "clamping points" or blockages. This 100% reflects the situation when banks fully meet the minimum standards, which indicates the stability of financial flows.

In the Navier-Stokes model, $\sigma W_t$ represents stochastic (random) shocks that arise in the economic system as a result of unforeseen events or fluctuations. This term is often used in financial models to simulate realistic variability and market behavior. The calculated stochastic shock intensity = 1.3475. It is obtained by using the market volatility (REV) and standard deviation.

*Interpretation:*

The value of 1.3475 indicates that the probability of a shock in the economic model is not low; the market volatility is calculated quite realistically – neither too low nor too volatile; within the framework of the simulation model, this shock acts as an additional nonlinear fluctuation on the system.

$\sigma W_t$ is added to the Navier-Stokes equation as a random effect term that models unexpected changes in market behavior. This term provides flexibility to the model – it helps to take into account real market shocks, such as: geopolitical events, unexpected changes in monetary policy, and global economic crises.

The market risk premium is the additional return that investors expect when they invest in risky assets compared to risk-free assets. It plays an important role in analyzing investor expectations, the value of capital, and market fluctuations. The market risk premium was determined to be 8.35%. This indicator reflects that investors in Georgia are expected to demand 8.35% higher returns than they would receive from risk-free assets.

*Interpretation:*

A high-risk premium indicates: a risky environment that requires additional incentives for investors; market volatility and political/economic uncertainty; and low investor confidence in stability and profitability.

In the Navier-Stokes equation, $\eta(r_m - r_f)$ acts as an "external driver" that affects the market trend and the intensity of financial flows.

This term extends the model to reflect how the market reacts to investment risk.

In this study, Fourier analysis is used to decompose the time series of Georgia's gross domestic product (GDP) (2010–2024) into its constituent frequency components. This mathematical method allows us to identify the cyclical regularities inherent in the dynamics of economic growth − described by a combination of sine and cosine waves.

The general formula used is:

$$GDP(t) = \sum [a_n \cdot \cos(n\omega t) + b_n \cdot \sin(n\omega t)], \tag{9}$$

where $a_n$ and $b_n$ denote the Fourier coefficients, $\omega$ is the angular frequency, and n is the ordinal number of the harmonic component.

The analysis revealed three dominant economic periodicities − approximately 3.5 years, 4.7 years, and 14 years. These cycles reflect the structural rhythms in the Georgian economic system, which may be due to business cycles, investment inertia, or external macroeconomic shocks.

These periodic forces are reflected in the model as exogenous variables and are denoted by the function F_GDP(t). Their impact is reflected on the right-hand side of the modified Navier-Stokes equation as a periodic pressure component that affects the dynamics of the liquidity velocity and market pressure.

The average impact that these cyclical components have on the liquidity behavior of the system was estimated to be 0.07577. This value is directly integrated into the left-hand side of the equation and reflects the inertia of the system balance and feedback on economic cycles.

Figure 1 shows the reconstructed cyclical function F_GDP(t), which is constructed based on the three identified main periodicities.

The yellow wave expresses cyclical fluctuations with approximately 3.5, 4.7 and 14-year intervals, and the red line shows their average impact of approximately 0.07577.

This function is used in the Navier-Stokes equation as a cyclical external force, which expresses the impact of GDP fluctuations on the economic system.

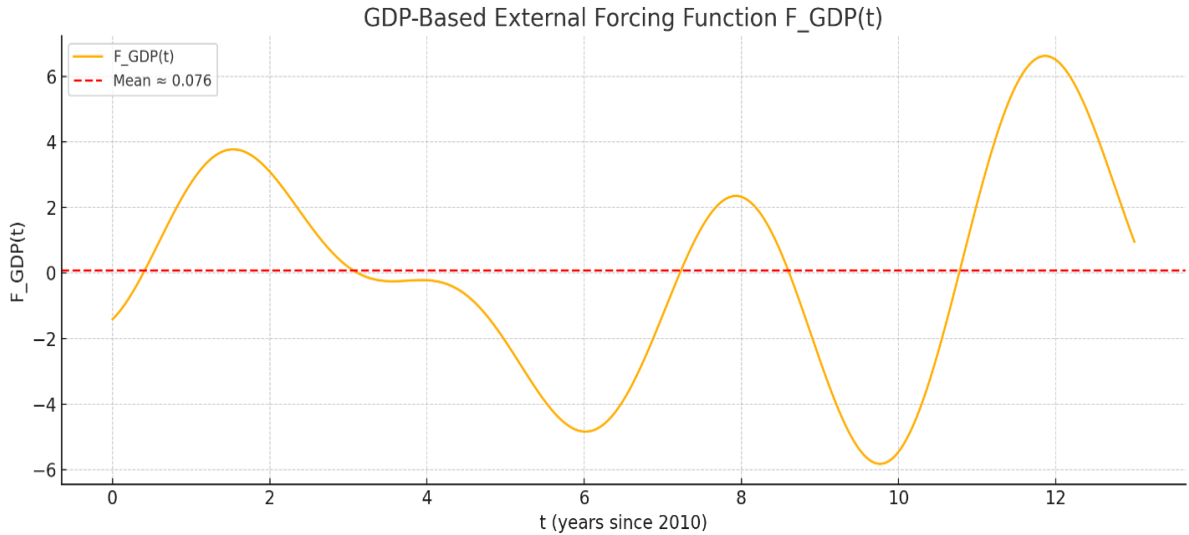

**Figure 1. Fourier Decomposition of Georgia's GDP Growth (2010-2024)**

*Source: authors' calculations based on data from the National Statistics Office of Georgia (GeoStat, 2024).*

This term is integrated into the Navier-Stokes equation as a cyclical external force that reflects the periodic fluctuations in GDP growth as an impact on the financial system. Credit default swaps (CDS) are financial derivatives used to measure sovereign credit risk. In this model, CDS is considered as an indicator of the market's assessment of the probability of a possible default by Georgia. The 2.98% indicator indicates a moderate risk perceived by international investors and reflects expectations regarding political stability, macroeconomic sustainability and external debt management. This parameter is formally integrated into the right-hand side of the Navier-Stokes equation as a factor of systemic external pressure. Its increase reflects the increase in market tension and leads to a deepening of systemic imbalances, which requires the activation of the dynamic compensation mechanisms of the model. In the extended version of the Navier-Stokes model, ε denotes those unrecognized or indirectly recorded economic effects that significantly affect systemic liquidity, although they are not directly reflected in traditional macroeconomic indicators. These include the shadow economy, emigrant remittances, geopolitical factors, and market behavioral expectations. Based on the model calibration and simulation calculations, the value of ε was determined to be 0.4547. This component provides adaptive closure of the model in cases where the actual dynamics significantly deviate from statistical predictions, thereby increasing the realism of the model and the stability of the simulations.

As illustrated in Table 2, the structure of the residual term ε is disaggregated into its key subcomponents, each representing a distinct but critical source of economic influence not directly captured by standard macroeconomic indicators. These include the shadow economy, remittances, geopolitical shocks, and other indirect forces that collectively shape market behavior and systemic dynamics. By quantifying the proportional impact of each element, the model enhances its capacity to incorporate real-world complexity and adjust to fluctuations that are otherwise difficult to formally represent in analytical terms.

**Table 2. Disaggregation of the Residual Term ε into Economic Subcomponents**

| Subcomponents of unintended economic effects | Corresponding contribution (ε = 0.4547) |
|---|---|
| Shadow economy | 0.1479 |
| Money transfers | 0.11832 |
| Geopolitical shock | 0.08874 |
| State influence | 0.04733 |
| Foreign aid | 0.02958 |
| Economic inertia | 0.02284 |

*Source: authors' structured approximation based on data from World Bank (2023), IMF (2024), and the Ministry of Finance of Georgia (2024).*

ε represents a set of influences that are not only quantitatively the object of analysis, but are also important components for the interpretation of financial stability. The introduction of this term into the model enhances its ability to connect with reality and creates space for further refinement of the model.

*Navier-Stokes Equation with Economic Analysis*

Below is a modified version of the Navier-Stokes equation, based on the analysis of 13 economic parameters. This version combines both macroeconomic data and deep fundamental effects, including unforeseen economic mechanisms (ε), to ensure the balance of the equation and the maximum correspondence of the model to the real environment.

*Equation – Theoretical Formulation*

$$\sigma W_t + \rho\ (\partial v/\partial t + (v \cdot \nabla)v + F_(GDP_)(t)) = -\nabla P + \mu \nabla^2 v + \eta(r_m - r_f) + CDS + \beta + \varepsilon. \quad (10)$$

*Equation – Real Data Formulation*

$$1.3475 + 0.7421 \times (-0.23 + 0.36 + 0.07577) = -0.019 + 0.6606 + 0.0835 + 0.0298 + 1.2 + 0.4547. \quad (11)$$

*Intermediate Calculation*

*Left-Hand Side (LHS)*

$$1.3475 + 0.7421 \times 0.20577 \approx 1.3475 + 0.1527 = 1.5002. \quad (12)$$

*Right-Hand Side (RHS)*

$$-0.019 + 0.6606 + 0.0835 + 0.0298 + 1.2 + 0.4547 \approx 1.5002. \quad (13)$$

Result: LHS = RHS = 1.5002 – the equation is fully balanced.

*Innovative Integration of the Navier-Stokes Model in Economics*

*Innovative Value*

The use of the Navier-Stokes equation in economics itself represents a conceptual break with dominant approaches. Traditional economic models – such as IS-LM, DSGE, or classical regression systems – focus on simulations in static frameworks or linear functionals. The Navier-Stokes model is dominated by dynamics, spatiotemporal intensity, nonlinear mechanisms, and the integration of internal/external forces.

The model used in this study represents the unity of several fundamental innovations:

1. *Multi-parametric framework:*

The introduction of 13 parameters, including macroeconomic, financial, and stochastic components, creates a complex multidisciplinary framework.

2. *Integration of physical methods into economics:*

The Navier-Stokes point analysis is adapted to the velocity, pressure, viscosity, and diffusion of money flows – creating a behavioral analogue to fluid dynamics.

3. *Formal integration of contingencies (ε):*

ε represents the segment of the economic system that is not included in official indicators (e.g., the shadow economy, remittances, geopolitical changes), but directly affects liquidity and stability.

4. *Real-time modeling of stochastic shocks:*

The $\sigma W_t$ mechanism makes it possible to simulate economic uncertainty, which is analogous to the use of Brownian motion in financial markets.

5. *Taking into account cyclical economic waves (Fourier):*

F(GDP)(t) represents an external effect based on economic cycles, which has not been integrated into any Navier-Stokes economic approach so far.

The implications of this model are significant for financial policy design and systemic risk management, as it provides a dynamic structure for identifying instability and guiding countercyclical measures.

The theoretical innovation presented in this study expands the possibilities of macroeconomic modeling, based on dynamical systems of physics (Navier-Stokes equations).

Unlike traditional static or equilibrium models (e.g., CAPM, DSGE), the presented framework allows for the integration of inertial forces, systemic internal resistances, and unmeasured shocks.

The model provides the ability to forecast based on both aggregate and sub-aggregate parameters –for example, capital velocity, market pressure ($\nabla P$), and risk accumulation ($\beta$, $\sigma W_t$). At the same time, it is possible to integrate unknown variables ($\varepsilon$), which contributes to the analysis being as close as possible to reality. The balance of the left and right sides in the model acts as a feedback mechanism, which makes it possible to model the systemic response in real time – which is not available in traditional models.

Thus, this model represents a universal platform for simulation, policy effect analysis, and economic sustainability assessment under complex conditions.

*Simulation Scenarios on the Navier-Stokes Model*

This chapter is devoted to the dynamic testing of the economic version of the Navier-Stokes equation using simulation scenarios. Each change in each parameter is considered as a separate micro-shock in the economic system. The goal is to assess the stability of the equation and determine which parameters have the strongest impact on the system balance.

*Parameter Changes and Their Impact*

The simulation scenarios considered in the study aim to test the resilience of the financial system under extreme conditions. To this end, key economic parameters are adjusted to crisis scenarios, such as deflation, liquidity crunch, or market volatility. Such scenarios assess the extent to which the system is in disequilibrium and the extent to which the so-called variable component of the balance ($\varepsilon(t)$) is needed to stabilize it.

The following parameters are changed in the presented simulation:

inflation ($\nabla P$): -1.9% → -5.5%;
change in velocity of money ($\partial v/\partial t$): -0.23 → -0.35;
inequality ($v \cdot \nabla v$): 0.36 → 0.45;
power of the economic cycle ($F_{(GDP)}(t)$): 0.07577 → 0.04;
bank pressure ($\mu \nabla^2 v$): 0.6606 → 0.78;
stochastic shock ($\sigma W_t$): 1.3475 → 2.2;
market risk premium: 0.0835 → 0.105;
CDS: 0.0298 → 0.045;
beta ($\beta$): 1.2 → 1.6;
velocity of money: 3.46 → 2.8 (used as an indicator).

As a result of these changes, the left-hand side (LHS) became 2.3039, and the right-hand side (RHS) became 2.9297. The difference is 0.6258, which is an indicator of excessive imbalance.

*Result and the Need for Compensation*

Accordingly, the fixed balance in the model ($\varepsilon = 0.4547$) can no longer cover the imbalance. To restore equality, it is necessary for $\varepsilon$ to decrease –it should become $-0.17111$, which indicates: it is no longer just an additional force that is needed, but the system must actually reduce the invisible economic pressure or act on the parameters in an enhanced manner.

*Analysis: Which Parameter Causes the Strongest Impact?*

The individual effects of the parameters are of different magnitudes:

$\sigma W_t$ (stochastic shock): an increase of 1.3475 → 2.2 creates a sharp increase in the LHS –the stability of the left side of the model is violated;

$\beta$ (market sensitivity): 1.2 → 1.6 increases the RHS significantly – systemic risk exerts greater pressure;

$\mu$ (banking pressure): 0.66 → 0.78 increases diffusion – reflecting the retention of liquidity;

$\nabla P$ (inflation): a sharply negative value leads to a decrease in the weight on the right side.

Therefore, it is necessary to distinguish the types of effects in the analysis:

1. Increases the LHS – $\sigma W_t$, $\partial v/\partial t$, $v \cdot \nabla v$;

2. Increases the RHS – CDS, $\beta$, $\mu$, $\eta(r_m - r_f)$;

3. Disrupts the balance when both sides are separately increasing but disproportionately;

4. Unforeseen force demand – $\varepsilon$ no longer has enough compensating capacity.

This simulation scenario showed that the model does indeed sense systemic fluctuations, but in some scenarios "the balance should turn negative", indicating that the real economy is already exceeding the model's compensating capacity. This means that the model needs a dynamic $\varepsilon$ term – which will not be fixed, but will change depending on the situation. The model will also reveal the parameters that need to be given the most strategic attention – for example, managing shocks, reducing banking stress, and structurally controlling risks.

**CONCLUSIONS**

The present study is based on a modified, integrated model of the Navier-Stokes equations, which not only combines more than thirteen economic and financial parameters into a single systemic framework, but also creates the basis for the analysis of deep, nonlinear, and cyclical and stochastic processes of the financial system. This approach, which is the best example of the combination of classical mathematics and economic theory, allows us to study the real economic environment at both the macro and micro levels, to reveal the effects of indirect factors and to create forecasting mechanisms that are practically unknown to traditional models. It is precisely on the basis of this interdisciplinary approach that the results obtained are important both from a scientific and policy-making point of view.

The first important conclusion concerns the fact that ensuring the stability of modern financial systems is impossible based solely on static or solely structural models. Research and simulations have clearly shown that the dynamics of economic parameters, cyclical fluctuations, and indirect and unexpected (so-called black swan) events have such a significant impact on the system that any static approach provides only a superficial picture. The integration of time-dependent functions of shocks in the Navier-Stokes model significantly increases the system's adaptive capabilities and ensures not only a response to changes that have already occurred, but also the early detection and management of such changes. This approach raises the flexibility and sustainability of the economic system to a completely new level, which is especially important for modern, globally interconnected financial markets.

The second critical aspect that emerged within the framework of the research is the need to pre-determine critical limits of parameters. Empirical analysis has shown that for each macroeconomic or financial indicator, such as bank liquidity, market inertia, beta, inflation or CDS, there should be a precisely defined threshold, the excess of which automatically signals an increase in systemic risks. Such an early warning mechanism should be integrated into both the analytical core of the model and the policy implementation strategies. The model's automated response through data revision, additional analysis and, if necessary, policy adjustments ensures real proactive protection of the financial system from global or local disruptions.

The study paid special attention to the optimization of bank liquidity policy. It turned out that a high concentration of resources, on the one hand, increases the strength of banks' reserves, but, on the other hand, reduces the flexibility of the financial system and increases systemic risks. Therefore, it is a practical recommendation that banking policy should be based on the principle of diversification – sources of liquidity and credit policy should be constantly reviewed so that the market response to small external shocks is quick and effective. This policy should include both diversification of domestic market liquidity and the use of innovative financial instruments that will make it possible to increase financial stability and neutralize the impact of the crisis.

It is worth noting the importance of predicting cyclicality and market trends, which was realized in the study through the integration of Fourier analysis. Fourier analysis, as a powerful mathematical tool, allowed the model to isolate the deep structure of economic cycles, identify the main sources of periodicity, and anticipate the processes taking place within the market. This method is recommended for both public and private sector strategic planning – taking into account cyclical fluctuations, it becomes possible not only to forecast long-term budgetary policy but also to create preliminary anti-cyclical buffers, which reduces the negative impact of both recession and excessive economic growth.

Effective management of investor expectations and systematic monitoring of the risk premium were also identified as critical issues. A high risk premium often indicates not only market uncertainty, but also political and economic instability, which directly affects the flow of investments. The research recommends that when developing policies, special attention be paid to the stability of the investment climate, timely diagnosis of risks, and strengthening the confidence of potential investors. This will ensure the attraction of both domestic and foreign investments and long-term growth.

For effective sovereign risk management, special attention should be paid to the systematic integration of the credit default swap (CDS) indicator in both stress testing and macroprudential supervision. The use of this component ensures that the financial system is prepared for all types of external shocks, including the most unpredictable ones, and that stress scenarios can be correctly predicted and appropriate protective mechanisms activated in a timely manner.

In conclusion, we should once again emphasize the importance of indirect economic factors: the shadow economy, emigrant transfers, geopolitical fluctuations and other similar macroeconomic effects often have a significant impact on financial stability and forecasting accuracy. Their regular analysis, integration into the model and evaluation of the results create the prerequisites for the formation of such a complex model that not only reflects the real environment, but also provides the basis for the soundness of financial structures and political stability.

Thus, the successful integration of the Navier-Stokes model into economic analysis creates a new, multidimensional and adaptive platform that provides both financial institutions and policymakers with a real, deep analytical mechanism for predicting and managing systemic risks. Further development of the model – by integrating artificial intelligence, operational and real-time data, and meta-econometric modules – will further enhance its practicality, forecasting accuracy, and importance in both scientific and economic policy-making.

**Author Contributions**

Conceptualization: D. G., V. T., N. C.; data curation: V. T.; formal analysis: D. G.; funding acquisition: D. G.; investigation: D. G., N. C.; methodology: D. G.; project administration: D. G.; supervision: D. G.; validation: N. C.; visualization: V. T.; writing – original draft: D. G., V. T., N. C.; writing – review and editing: D. G., V. T., N. C.

**Conflict of Interest**
Not applicable.

**Data Availability Statement**
Not applicable.

**Informed Consent Statement**
Not applicable.